# Topology Optimization of Turbulent Fluid Flow with a *Sensitive Porosity Adjoint Method* (SPAM)

B. Philippi and Y. Jin[*]

Thermo-Fluid Dynamics, Hamburg University of Technology, 21073 Hamburg, Germany

* Corresponding author, tel: +49(0)40 428782676, fax: +49(0)40 428784169

e-mail: yan.jin@tuhh.de; jinyan995072@gmail.com

**Abstract**

A *sensitive porosity adjoint method* (SPAM) for optimizing the topology of fluid machines has been proposed. A sensitivity function with respect to the porosity has been developed. In the first step of the optimization process, porous media are introduced into the flow regime according to the sensitivity function. Then the optimized porous media are transformed to solid walls. The turbulent flow in porous media is accounted for by a modified eddy-viscosity based turbulence model. Its influence on the adjoint equations is nevertheless neglected, which refers to the so called *frozen turbulence assumption*. A test case of application in terms of the turbulent rough wall channel flow shows that a considerable reduction of the objective function can be obtained by this method. The transformation from porous media to solid walls may have important effect on the optimization results.

## 1 Introduction

Shape/topology optimization of fluid machines is an important task in many industries, e.g. aerospace, power generation and chemical engineering. With the optimized shape, one may expect to increase the lift and reduce the drag for an airfoil, enhance the heat transfer for a heat exchanger, increase the yield of a chemical reactor, etc.

However, shape/topology optimization for a flow problem is also a challenging task since a large amount of variables need to be taken into account. When the optimization method is coupled with a *computational fluid dynamic* (CFD) method, the number of the variables to be optimized has the same order as the facet number at the boundary surfaces or the cell number in the flow regime.

Adjoint methods, which are based on the method of *Lagrange multiplier* [1] have received increasingly more interest in recent years. This approach may yield optimized design without using too much computing resources. Another advantage of using adjoint methods in fluid dynamics is the free choice of the cost function, which can be formulated for either volume or surface related objectives.

Adjoint methods can be applied to both shape and topology optimization problems. The former one deals with the modification of wall surfaces. In an early research, Giles and Pierce [2] introduced how to use the adjoint method to optimize the design of business jets. More details with respect to this method for shape optimization are discussed in [3]. Jameson [4] reviewed the formation and application of the optimization technique based on the adjoint method for aerodynamic design. Löhner and Soto [5, 6] developed an optimization approach which couples the adjoint method with the CFD technique. All points on the surface can be treated as design parameters and optimized by this approach. Zymaris et al. [7] developed an adjoint shape optimization method for turbulent flows. The effect of turbulence models and wall functions on the adjoint equations has been considered in this method. Stück and Rung [8] implemented the adjoint RANS method in the framework of an unstructured finite volume code. An explicit filtering technique is introduced to remove the numerical noise.

For a topology optimization, the geometry is not described by the surface parameters but with the volume elements in the entire domain. This characteristic makes the optimization more flexible since more geometrical variables can be controlled and optimized. The adjoint method for topology optimization has received intensive investigations since the early work by Borrvall and Petersson [9]. Gerborg-Hansen et al. [10] demonstrated how to optimize laminar channel flows with this method. Based on this method, Guest and Prevost [11] developed a so called *Darcy-Stokes approach* to optimize the creeping fluid flows. Olesen et al. [12] coupled this method with the commercial CFD software package FEMLAB. The developed optimization solver was tested by studying steady state viscous flows. Srinath and Mittal [13] proposed a *stabilized element formation* to solve the adjoint equations. The solver is validated by calculating the flow past an elliptical bump. Othmer [14, 15] has further improved this method and implemented it to the CFD code OpenFOAM. Hinterberger and Olesen [16] optimized the automotive exhaust systems by this method.

The topology adjoint methods mentioned above always employ a punishing term in the momentum equations to approximate porous medium flows. The Forchheimer term and turbulence effect are nevertheless neglected in these methods. In addition, the effect of transformation from porous media to real solid walls hasn't received much attention. Under these considerations, a so called *sensitive porosity adjoint method* (SPAM) has been proposed in the present study. The approach is developed based on the previous work, especially [14] and [15]. Compared with the previous work, the following improvements have been made:

- A complete porous media model including both the Darcy's term and the Forchheimer's term has been considered.
- The effect porous medium on the turbulent flow has been taken into account.
- The sensitivity function with respect to the porosity has been proposed.
- The effect of transformation from porous media to solid walls has been analyzed.

## 2 The optimization problem

Here we consider an optimization problem of minimizing the objective function *J*. The objective function in the present study is considered to be only related to the boundary values, i.e.

$$J = \int_{\Gamma} J_{\Gamma} dA \tag{1}$$

The constraint condition is the Reynolds averaged Navier-Stokes equations for incompressible turbulent flow in a porous medium, they are

$$R_i^v = \frac{\partial (v_i v_j)}{\partial x_j} + \frac{\partial p}{\partial x_i} - \frac{\partial}{\partial x_j}\left(v_{eff} \frac{\partial v_i}{\partial x_j}\right) + \alpha v_i + \beta |v_i| v_i = 0 \tag{2}$$

$$R^p = -\frac{\partial v_i}{\partial x_i} = 0 \tag{3}$$

where $\alpha$ and $\beta$ are the coefficients in the Darcy's term and the Forchheimer's term. They are calculated by

$$\alpha = \frac{v}{K}; \quad \beta = \frac{C_F}{K^{1/2}} \tag{4}$$

The coefficient $C_F$ is set to be a constant 0.55 since the element size of the porous media $d_p$ is assumed to be much smaller than the macroscopic length scale. The permeability *K* is calculated by the Carman-Kozeny's equation [17, 18]:

$$K = \frac{d_p^2 \phi^3}{180(1-\phi)^2} \tag{5}$$

where $\phi$ is the porosity of the porous media.

2. 1 Turbulence treatment

The eddy viscosity assumption is often employed for clear turbulent flows ($\phi = 1$) in order to close the Reynolds averaged Navier-Stokes equations. Under this assumption, the so called eddy viscosity $v_t$ is introduced. There are different models for calculating the value of $v_t$ such as k-ε family models [19-21] and k-ω family models [22, 23].

For a porous medium flow, The commercial CFD package FLUENT 13 [24] suggests to assume that the solid medium has no effect on the turbulence transportation. This assumption may be reasonable when the medium's porosity $\phi \to 1$ since the geometric scale of the medium does not interact with the scale of the turbulent eddies. However, the condition of $\phi \to 0$ is important to the present optimization method and porous medium effect at this condition must be considered.

In the recent study, Jin et al. [25] has found that the largest scale of the turbulent structures in porous media are proportional to the pore size, which is determined by the porosity $\phi$. According to this trend, we tentatively specify the eddy viscosity in a porous medium by

$$v_{tp} = \phi \cdot v_t \tag{6}$$

where $v_t$ is calculated with the turbulence models for clear flows. With this assumption, all the available turbulence models for clear flows can be easily employed by the present method without modification of their transportation equations.

Some turbulence models [26-28] emerge in recent years, in which the effect of porous media on the transportation equations of turbulence variables such as $k$ and $\varepsilon$ are accounted for. The present model is identical to these models when $\phi \to 1$ and 0, which are the regions of our interest in this study. Thus, the effective viscosity in Eq. (2) is calculated by

$$v_{eff} = v + v_{tp} = v + \phi v_t \tag{7}$$

2.2 Boundary conditions

Three often used boundary conditions are discussed here:

$$\text{Inlet: } v_{t1} = v_{t2} = 0, \; v_n = v_{n0}, \; \partial p / \partial x_i \cdot n_i = 0 \tag{8}$$

$$\text{Outlet: } p = p_0, \quad \partial v_i / \partial x_j \cdot n_j = 0 \tag{9}$$

$$\text{Wall: } v_i = 0, \quad \partial p / \partial x_i \cdot n_i = 0 \tag{10}$$

where $n_i$ denotes the direction vector of a boundary surface. The subscripts $n$, $t1$ and $t2$ denote the normal and (two) tangential direction components.

Thus, the optimization problem of the present study is to minimize the objective function $J$, which subject to the governing equations (1)-(2) and the boundary conditions (5)-(7).

## 3. Sensitive porosity adjoint method (SPAM)

### 3.1 Lagrange function

To minimize the objective function $J$, a Lagrange function $L$ is constructed according to the method of *Lagrange multipliers* [1], i.e.

$$L = J + \int_\Omega \left( R_i^v \cdot u_i + R^p \cdot q \right) dV \tag{11}$$

where $u_i$ and $q$ are the adjoint velocity component and pressure respectively. The existence of an optimized $J$ requests

$$\frac{\partial L}{\partial b_m} = 0 \tag{12}$$

Here it is assumed that the transportation of the turbulence variables have little effect on the optimization method, which refers to the so called *frozen turbulence assumption*. Then $b_m$ can be any cell value of $v_i$, $p$, $u_i$, $q$ and $\phi$. When $b_m$ is a cell value of $u_i$ or $q$, Eq. (12) may lead to the governing equations (2)-(3). The adjoint equations may be derived when $b_m$ is equal to the a cell value of $v_i$ or $p$. The sensitivity function with respect to the porosity $\phi$ is derived when $b_m$ is equal to a cell value of $\phi$.

### 3.2 Adjoint equations

#### 3.2.1 Governing adjoint equations

Here we consider the objective function $J$ is a boundary value. When $b_m$ is any cell value of $v_i$ or $p$, the operator $\partial/\partial b_m$ can be moved into the integration operator. Thus Eq. (12) becomes

$$\frac{\partial}{\partial b_m} \int_\Gamma J_\Gamma dA + \int_\Omega \frac{\partial}{\partial b_m} \left( u_i R_i^v + q R^P \right) dV = 0 \tag{13}$$

The components in Eq. (13) can be rearranged as follows:

$$\frac{\partial}{\partial b_m}\left(u_i \frac{\partial (v_i v_j)}{\partial x_j}\right) = \frac{\partial}{\partial b_m}\frac{\partial (u_i v_i v_j)}{\partial x_j} - \frac{\partial v_i}{\partial b_m}\cdot\left(v_j \frac{\partial u_i}{x_j}\right) - v_i \frac{\partial v_j}{\partial b_m}\frac{\partial u_i}{x_j} \tag{14a}$$

$$\frac{\partial}{\partial b_m}\left(u_j \frac{\partial (p)}{\partial x_j}\right) = \frac{\partial}{\partial b_m}\frac{\partial (u_j p)}{\partial x_j} - \frac{\partial p}{\partial b_m}\cdot\frac{\partial u_j}{\partial x_j} \tag{14b}$$

$$\frac{\partial}{\partial b_m}\left(u_i \frac{\partial}{\partial x_j}\left(v_{eff}\frac{\partial v_i}{\partial x_j}\right)\right) = \frac{\partial}{\partial b_m}\frac{\partial}{\partial x_j}\left(v_{eff} u_i \frac{\partial v_i}{\partial x_j} - v_{eff} v_i \frac{\partial u_i}{\partial x_j}\right) + \frac{\partial v_i}{\partial b_m}\frac{\partial}{x_j}\left(v_{eff}\frac{\partial u_i}{\partial x_j}\right) \tag{14c}$$

$$\frac{\partial}{\partial b_m}\left(u_i(\alpha v_i + \beta|v_i|v_i)\right) = \frac{\partial v_i}{\partial b_m}\cdot(\alpha u_i + \beta|v_i|u_i) \tag{14d}$$

$$\frac{\partial}{\partial b_m}\left(q \frac{\partial v_j}{\partial x_j}\right) = \frac{\partial (q v_j)}{\partial b_m} - \frac{\partial v_j}{\partial b_m}\cdot\frac{\partial q}{\partial x_j} \tag{14f}$$

Substituting them into Eq. (13) and using the Gauss theorem, we have

$$\frac{\partial L}{\partial b_m} = \frac{\partial}{\partial b_m}\int_\Gamma J_\Gamma dA + \frac{\partial}{\partial b_m}\int_\Gamma \left(u_i v_i v_j + u_j p - v_{eff}\left(u_i \frac{\partial v_i}{\partial x_j} - v_i \frac{\partial u_i}{\partial x_j}\right) - q v_j\right) n_j dA$$

$$\boxed{- \int_\Omega \frac{\partial v_i}{\partial b_m}\cdot\left(v_j \frac{\partial u_i}{x_j} + v_j \frac{\partial u_j}{x_i} - \frac{\partial q}{\partial x_i} + \frac{\partial}{\partial x_j}\left(v_{eff}\frac{\partial u_i}{\partial x_j}\right) - (u_i \alpha + \beta|v_i|u_i)\right)dV}$$

$$\boxed{- \int_\Omega \frac{\partial p}{\partial b_m}\cdot\left(\frac{\partial u_j}{\partial x_j}\right)dV}$$

$$= 0 \tag{15}$$

The volume integration terms indicated in the frame lead to

$$\frac{\partial u_j}{\partial x_j} = 0 \tag{16}$$

$$v_j \frac{\partial u_i}{x_j} + v_j \frac{\partial u_j}{x_i} - \frac{\partial q}{\partial x_i} + \frac{\partial}{\partial x_j}\left(v_{eff}\frac{\partial u_i}{\partial x_j}\right) - (\alpha + \beta|v_i|)u_i = 0 \tag{17}$$

Eqs. (16) and (17) are the governing adjoint equations.

3.2.2 Boundary conditions

When the governing equations (16)-(17) are satisfied, Eq. (13) reduces to

$$\frac{\partial}{\partial b_m}\int_\Gamma J_\Gamma dA + \frac{\partial}{\partial b_m}\int_\Gamma \left(u_i v_i v_j + u_j p - v_{eff}\left(u_i \frac{\partial v_i}{\partial x_j} - v_i \frac{\partial u_i}{\partial x_j}\right) - qv_j\right) n_j dA = 0 \quad (18)$$

The boundary conditions of the adjoint equations is determined by Eq. (18). Here again we only discuss the three often used boundary conditions, they are inlet, outlet and wall conditions.

**Inlet:**

The effect of the diffusion term is considered to be much smaller than that of the convection term at inlet. Thus, the term $v_{eff}\left(u_i \frac{\partial v_i}{\partial x_j} - v_i \frac{\partial u_i}{\partial x_j}\right)$ in Eq. (18) is neglected to simplify the boundary conditions. Since the velocity components $v_i$ are fixed, the presser $p$ is the only variable determined by the internal cell values. When $b_m$ is a cell value of $p$, Eq. (18) becomes

$$\frac{\partial J_\Gamma}{\partial p} + u_i n_i = 0 \quad (19)$$

Eq. (19) may determine one of the velocity components at the inlet. The adjoint pressure $q$ is extrapolated from the internal values. The other two velocity components are set to be zero.

**Outlet:**

The diffusion term $v_{eff}\left(u_i \frac{\partial v_i}{\partial x_j} - v_i \frac{\partial u_i}{\partial x_j}\right)$ in Eq. (18) is neglected again to simplify the boundary condition. Since $b_m$ can be anyone of the velocity components, Eq. (18) can be satisfied only when

$$q = u_i v_i + u_n v_n + \frac{\partial J_\Gamma}{\partial v_n} \quad (20a)$$

and

$$0 = u_t v_t + \frac{\partial J_\Gamma}{\partial v_t} \quad (20b)$$

where $n$ and $t$ denote the normal and tangential directions at the boundary surface. The normal adjoint velocity component $u_n$ is extrapolated from the internal field.

**Wall:**

The boundary conditions of walls are similar to those of inlet. When $J_\Gamma$ is 0, Eq. (18) leads to

$$u_i = 0 \tag{21}$$

The adjoint pressure $q$ is extrapolated from the internal value, i.e.

$$\partial q / \partial x_i \cdot n_i = 0 \tag{22}$$

3.3 Sensitivity function

When $b_m = \phi$, Eq. (12) leads to the sensitivity function with respect to $\phi$, i.e.

$$\frac{\partial L}{\partial \phi} = v_i u_i (\frac{d\alpha}{d\phi} + |v_i|\frac{d\beta}{d\phi}) = 0 \tag{23}$$

When Eqs. (4) and (5) are considered, $\dfrac{d\alpha}{d\phi}$ and $\dfrac{d\beta}{d\phi}$ can be calculated by

$$\frac{d\alpha}{d\phi} = -\frac{180}{d_p^2}\left(2\cdot\frac{1-\phi}{\phi^3} + 3\cdot\frac{(1-\phi)^2}{\phi^4}\right) \tag{24}$$

$$\frac{d\beta}{d\phi} = -\frac{7.38}{d_p}\left(\frac{1}{\phi^{3/2}} + \frac{3}{2}\cdot\frac{1-\phi}{\phi^{5/2}}\right) \tag{25}$$

The sensitivity function $\dfrac{\partial L}{\partial \phi}$ reflects the variation speed of $L$ with respect to the $\phi$ value at each cell. As demonstrated in Fig. 1, the cell values of $\phi$ are updated by

$$\phi_i^n = \phi_i^{n-1} + h_i \frac{\partial L}{\partial \phi_i} \tag{26}$$

where $i$ and $n$ are the indices of the cell ID and time instant respectively. The local marching step $h_i$ is determined by $h_i = \kappa \cdot \partial L/\partial \phi_i$ where $\kappa$ is a negative constant. Thus a local minimum $L$ can be found by Eq. 26. Usually, the local minimum $J$ can be obtained altogether with the minimum $L$.

The porosity $\phi$ is updated every time step according to Eq. (26) until $L$ and $J$ almost stop changing. However, the optimized $\phi$ is a continuous field which varies between 0 and 1. A critical porosity $\phi_c$ is introduced here in order to transform the porous media to solid walls, i.e.

$$\phi_{opt} = \begin{cases} 1 & \phi \geq \phi_c \\ 0 & \phi < \phi_c \end{cases} \tag{27}$$

Then the optimized topology of the flow regime, which is only composed of fluid zone ($\phi = 1$) and solid zone ($\phi = 0$), can be obtained.

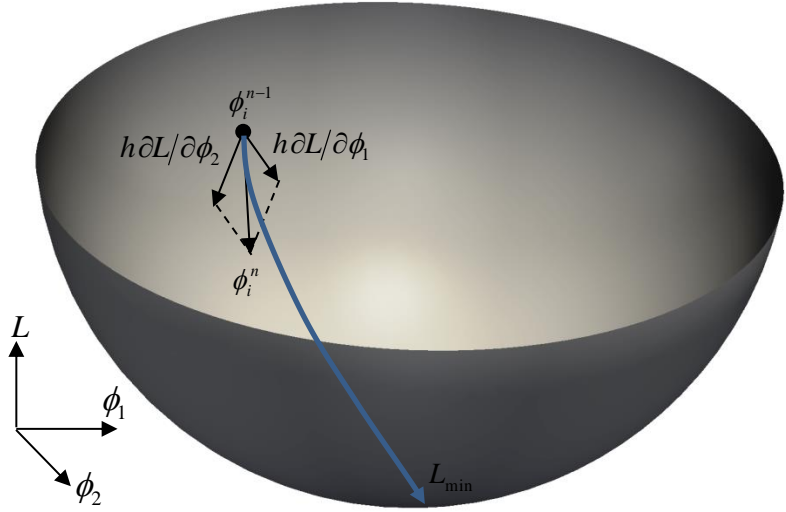

Figure 1: Schematic picture for updating $\phi_i$.

## 4. An example of application

4.1 Description of the test case

The test case is a two-dimensional channel with one wall mounted by 5 blade like obstacles, see Fig. 2. The computational parameters are given in table 1. In this test case we consider that these blades cannot be removed whereas additional solid structures can be introduced in the flow regime to optimize the flow. The objective of the test case is to minimize the loss of the mechanical energy. Thus $J$ is defined by

$$J = \int_\Gamma J_\Gamma dA = -\int_\Gamma \left(\frac{1}{2}v_i^2 + p + k\right)v_j n_j dA \tag{28}$$

Then the inlet boundary condition Eq. (19) becomes

$$v_i n_i + u_i n_i = 0 \tag{29}$$

The adjoint velocity in the surface normal direction $u_i n_i$ is determined by Eq. (29). The velocity component in the other direction is set to be zero. The outlet boundary condition Eq. (20) becomes

$$q = u_i v_i + u_n v_n - \frac{3}{2}v_n^2 \tag{30a}$$

and

$$u_t = v_t \tag{30b}$$

The other boundary conditions for the adjoint equations are specified according to section 3.2.2.

The optimization solver has been developed and implemented based on the open CFD source code OpenFOAM 2.3 [29]. The Navier-Stokes equations (2)~(3) and the corresponding ajoint equations (16)~(17) are solved by a steady state *finite volume method* (FVM). To compute the derivatives of the velocity, the variables at the interfaces of the grid cells are obtained with the 1$^{st}$ order bounded upwind scheme. The pressure $p$ and the adjoint pressure $q$ at the new time level are determined by the Poisson equations. The velocity $v_i$ and the ajoint velocity $u_i$ are corrected by the *SIMPLE method*.

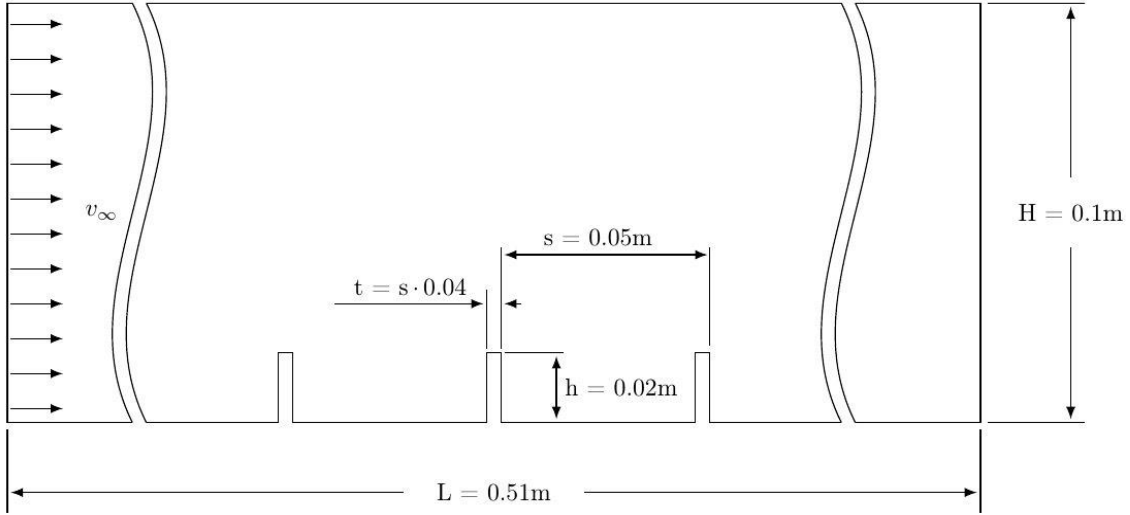

Figure 2 Geometry of the test case.

Table 1 Computational parameters

|  | unit | value |
| --- | --- | --- |
| inlet velocity $v_0$ | [ms$^{-1}$] | 1 |
| outlet pressure $p_0$ | [m$^2$s$^{-2}$] | 0 |
| kinematic viscosity $\nu$ | [m$^2$s$^{-1}$] | $10^{-6}$ |
| Re $= v_0 H/\nu$ | - | $10^5$ |
| mesh resolution | - | $1020 \times 200$ |

4.2 Flow fields before optimization

Fig. 3 shows the flow fields in the rough wall channel before optimization, including the distributions of porosity (Fig. 3a), streamlines (Fig. 3b), velocity magnitude (Fig. 3c), dissipation rate (Fig. 3d), turbulent kinetic energy (Fig. 3e) and turbulence viscosity (Fig. 3f). The original porosity is 1 since a clear flow is considered. The vortices in 6 zones divided by 5 blades are indicated in Fig. 3b. It can be seen that the interaction of vortices ② and ③ causes the instability of the flow.

Jin and Herwig [30, 31] showed that entropy and its generation rate are important for assessing a turbulent flow and heat transfer problem which is a typical irreversible process. In the present study, entropy generation is only due to dissipation rate $\varepsilon$ in the flow field since heat transfer is not considered. The dissipation rate is determined by

$$\varepsilon = 2\nu_{eff} s_{ij} s_{ij} \tag{31}$$

where the strain rate $s_{ij} = 1/2(\partial u_i/\partial x_j + \partial u_j/\partial x_i)$. Fig. 3d shows especially strong dissipation rate over the tips of the second and third blades, which correspond to the large vortex zones ② and ③. Similar phenomenon can be found in turbulent kinetic energy distributions, see Fig. 3e.

4.3 Optimized results

The turbulent flow is optimized by the method discussed in section 3. Figure 4 shows the optimized results before the porous media are transformed to solid walls. Some optimized porous media are connected to the wall surfaces of the channel while the others are suspended in the flow regime.

With the optimized structures, the objective function $J$ is reduced by 15%. The vortex sizes in zone ②, ③ and ⑥ are all reduced, see Fig. 4b. Both the dissipation rate $\varepsilon$ (see Fig. 4d) and the turbulent kinetic energy $k$ (see Fig. 4e) over the second and third blade are weakened.

Fig. 4a shows that the porosity is a continuous field which varies between 0 and 1. In an industrial application nevertheless it is more practical to optimize the flow with solid walls instead of porous media. Thus, the porous media are transformed to solid walls according to Eq. (27). Fig. 5 shows the reduction of $J$ for different $\phi_c$ values. For the present case, smaller $\phi_c$ values lead to higher $J$ reduction.

Fig. 6 shows the optimal results of $\phi_c = 0.1$. It can be seen in Fig. 6a that the optimized structures are made of a number of thin solid layers. The flow fields are similar to the porous medium results, except the slightly different vortical structures in zone ⑥, see Fig. 6b and 4b for comparison. It should be also noticed that the solid layers in zone ⑥ are not closed. There is a small hole which connects the external flow and the cavity. As a result, the dissipation, turbulent kinetic energy and turbulent viscosity are not zero in the cavity of zone ⑥.

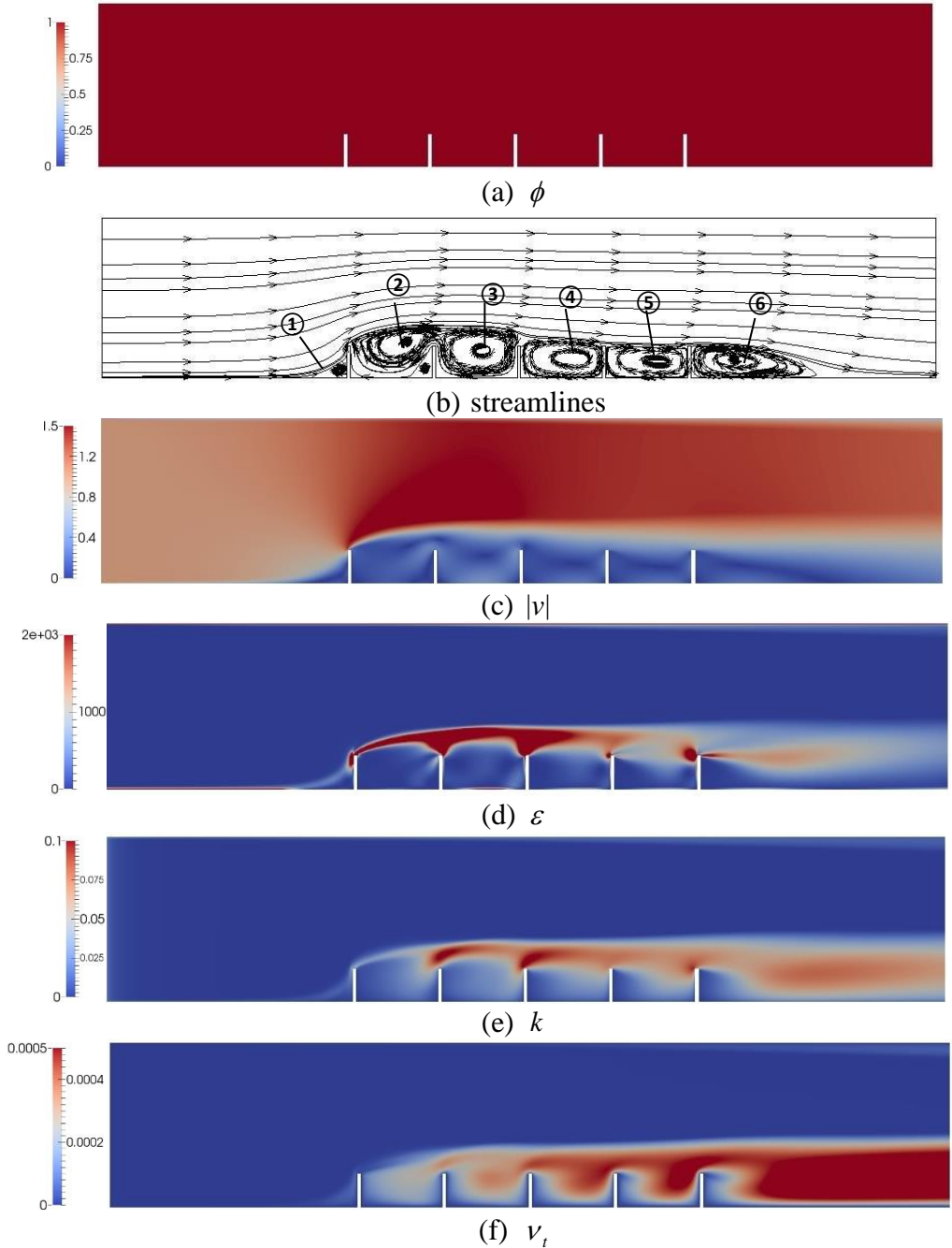

(a) $\phi$

(b) streamlines

(c) $|v|$

(d) $\varepsilon$

(e) $k$

(f) $v_t$

Figure 3 The orginal flow fields; $\phi = 1$; Re=$10^5$.

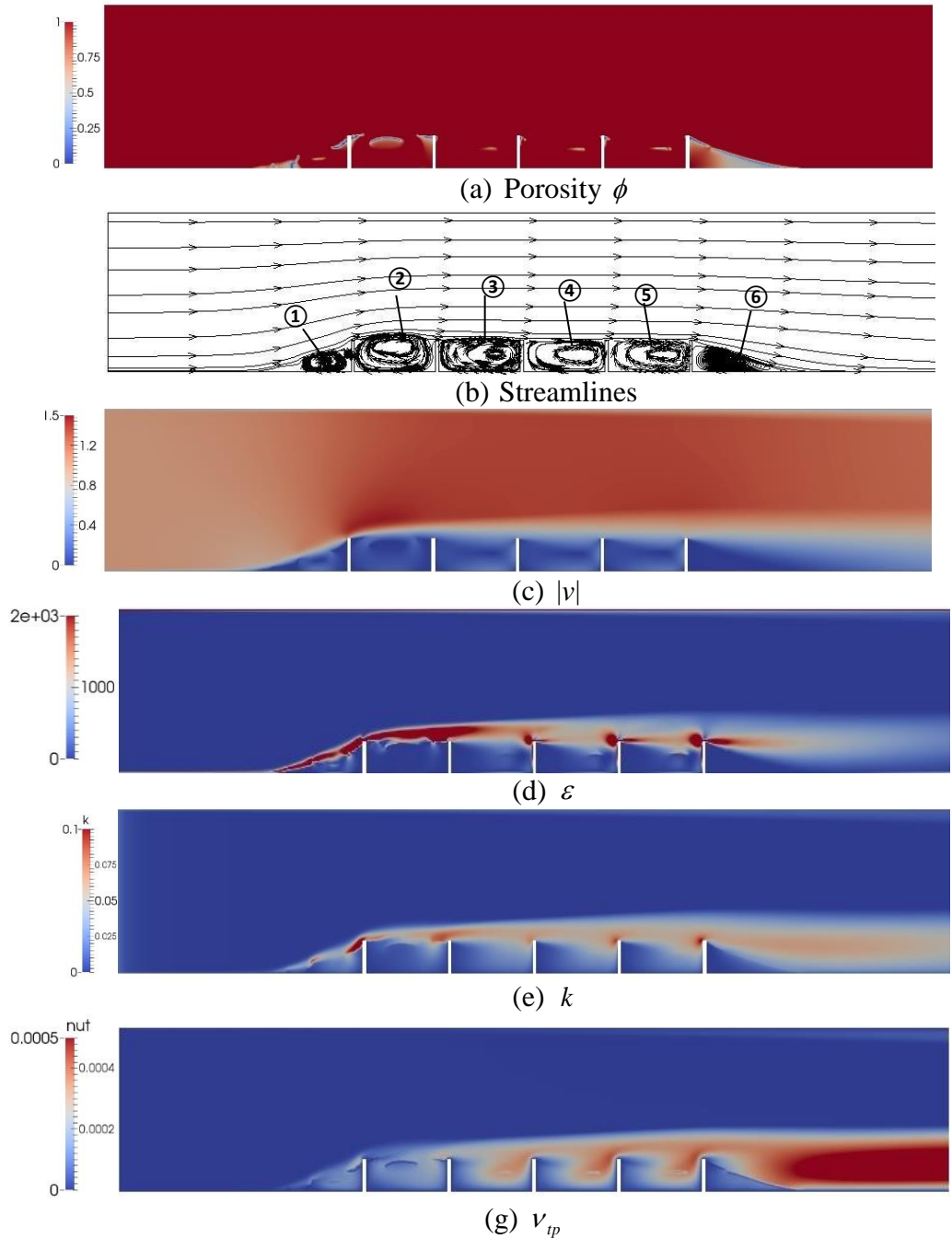

(a) Porosity $\phi$

(b) Streamlines

(c) $|v|$

(d) $\varepsilon$

(e) $k$

(g) $\nu_{tp}$

Figure 4 The optimized results; the porous media are not transferred to solid walls; Re=$10^5$; $(J-J_0)/J_0 = -15\%$; $J_0$ is the $J$ value of the original channel.

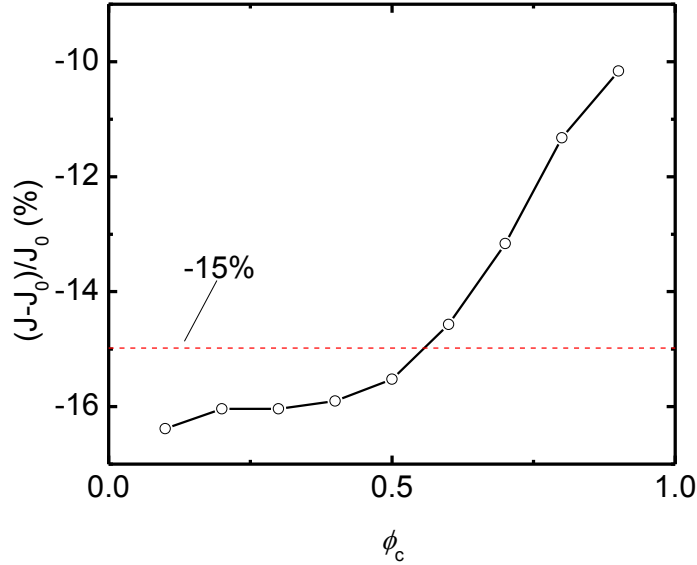

Figure 5 Reduction of $J$ when the porous media are transformed to solid walls;
------: porous medium optimization results; —o—: solid wall optimization results.

When $\phi_c = 0.9$, the optimized structures are made of solid blocks, see Fig. 7a. Only 10.2% reduction of $J$ is obtained by this structure. The solid structure in zone ⑥ doesn't have a hole connecting the cavity and the external flow. With this structure, the velocity magnitude, dissipation rate, turbulent kinetic energy and turbulent viscosity become close to zero in zone ⑥. However, the turbulent viscosity in the wake is much stronger than its value in case $\phi_c = 0.1$, see Fig. 7f and Fig. 6f for comparison. This difference is probably due to the reason that the cavity connected to the external flow may damp part of the turbulent fluctuations.

The solid structures in Fig. 6a can be further simplified to make them more feasible for applications. Fig. 8 shows four variations of design on the basis of Fig. 6a:

- Design A: The small structures in zones ③, ④ and ⑤ are removed;
- Design B: Both the small structures and the unconnected structures are removed;
- Design C: Only the solid structures in the wake (zone ⑥) are adopted;
- Design D: Only the solid structures upstream (zones ① and ②) are adopted.

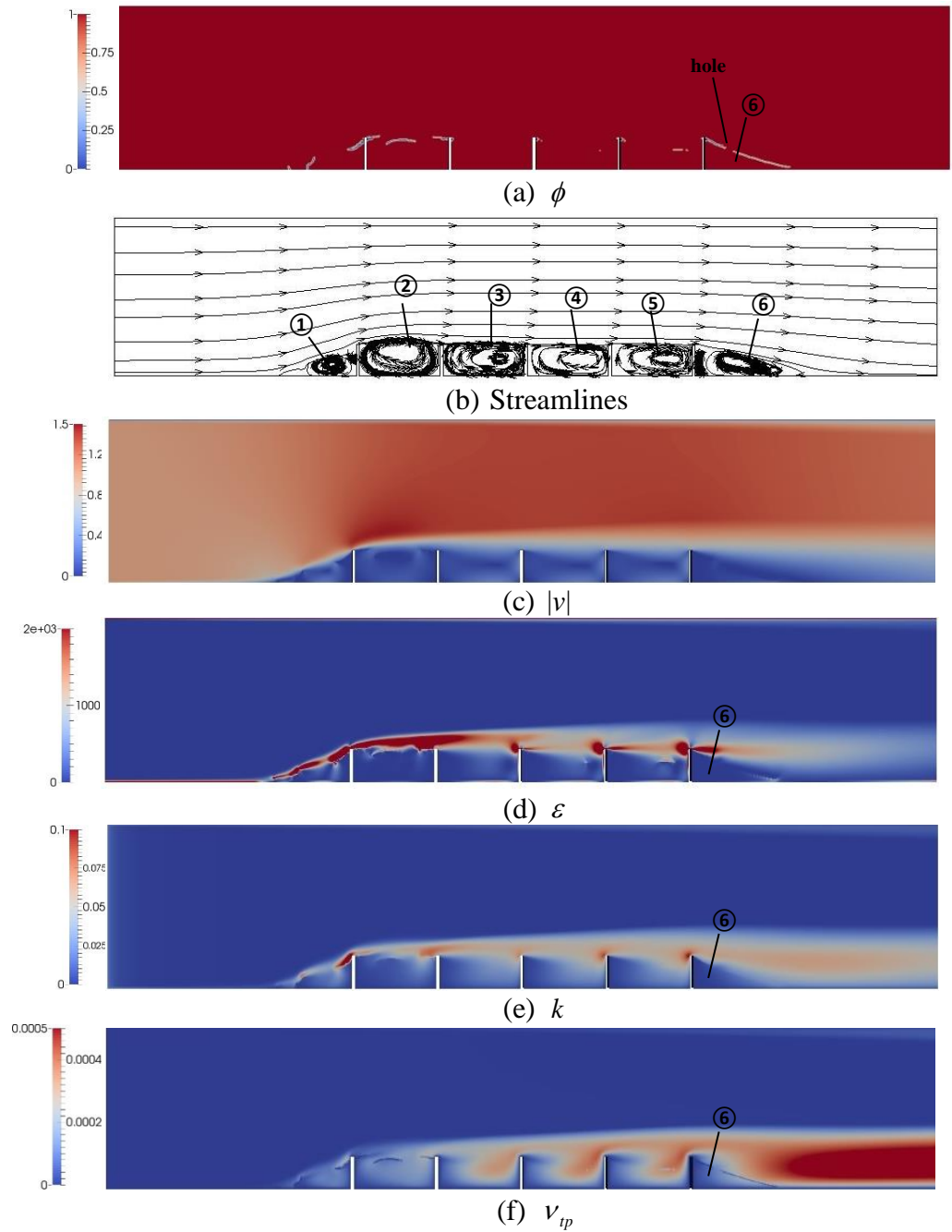

(a) $\phi$

(b) Streamlines

(c) $|v|$

(d) $\varepsilon$

(e) $k$

(f) $v_{tp}$

Figure 6 The optimized results; the porous media are transferred to solid with $\phi_c = 0.1$; Re=$10^5$; $(J - J_0)/J_0 = -16.4\%$.

The results show that the small structures in zones ③, ④ and ⑤ have almost no effect on the objective function *J*, thus they can be removed. The solid structures upstream have the most important contribution for reducing *J*. Design B has the most drag reduction, which suggests that the suspending structure in zone ② can be removed, see Fig. 8a and b for comparison.

4.4 Discussions

The test case shows that a typical Sensitive Porosity Adjoint Method (SPAM) is composed of two procedures:
- Optimizing the geometry by introducing porous media into the flow;
- Transforming the porous media into solid walls.

The present test case shows that the second procedure has a significant impact on the optimization results. The reduction of *J* varies between 10.2% and 16% when different critical porosity values $\phi_c$ are used.

The optimization results of the test case show that thin solid layers (smaller $\phi_c$) reduce more loss of mechanical energy than solid blocks (larger $\phi_c$). It is particularly interesting to observe that a hollowed cavity structure with a hole connected to the external flow may damp the turbulence in the wake, see zone ⑥ in Fig. 6 and 7 for comparison. However, the physics and generality of this structure need to be further investigated with more accurate numerical schemes and turbulence models.

The prerequisite of this optimization method is that the turbulent flow in porous media can be accurately calculated. However, this is a challenging task and further effort should be made to improve the turbulence models, especially for the turbulent flows with large scale vortical sheddings. At the current stage, despite of the model errors, we still expect an improved design with respect to fluid flows can be proposed by this method.

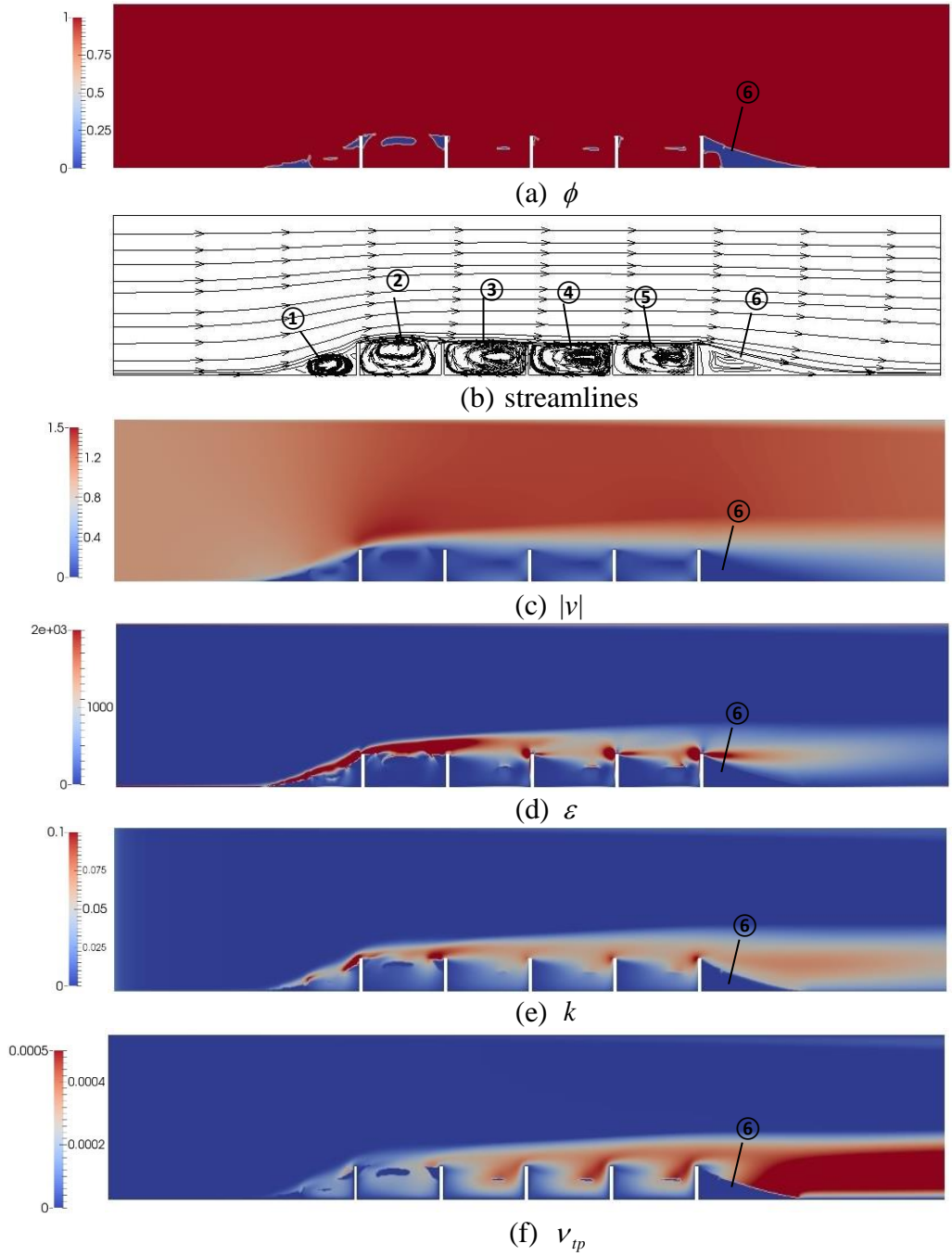

(a) $\phi$

(b) streamlines

(c) $|v|$

(d) $\varepsilon$

(e) $k$

(f) $\nu_{tp}$

Figure 7 The optimized results; the porous media are transferred to solid with $\phi_c = 0.9$; Re=$10^5$; $(J - J_0)/J_0 = -10.2\%$.

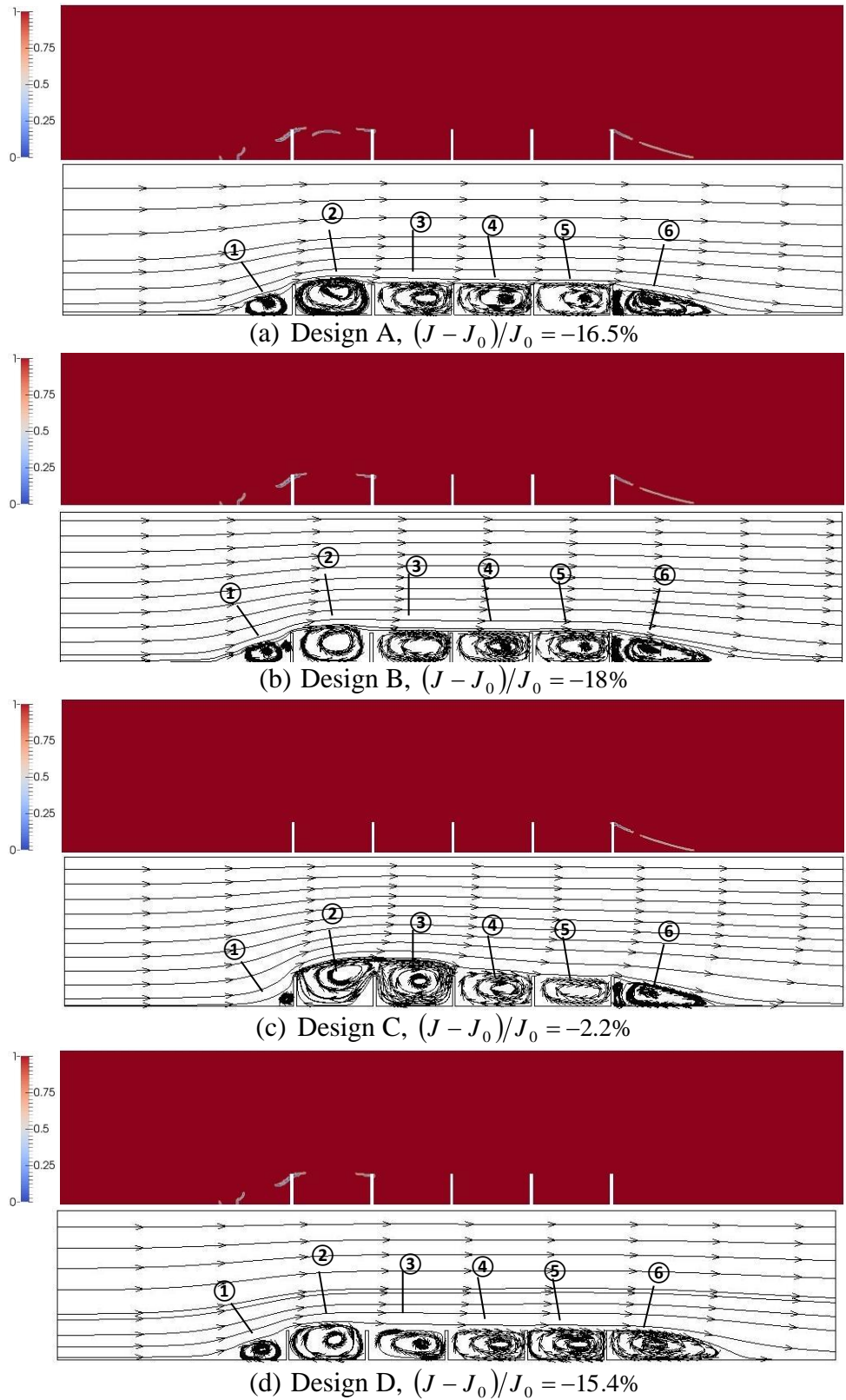

(a) Design A, $(J - J_0)/J_0 = -16.5\%$

(b) Design B, $(J - J_0)/J_0 = -18\%$

(c) Design C, $(J - J_0)/J_0 = -2.2\%$

(d) Design D, $(J - J_0)/J_0 = -15.4\%$

Figure 8 Variations of the design based on the solid strctures in Fig. 6a, Re=$10^5$. Here only the distribution of $\phi$ (upper figures) and the corresponding streamlines (lower figures) are shown.

## 5. Conclusions

A so called *sensitive porosity adjoint method* (SPAM) for optimizing the topology of the fluid machines has been developed in the present study. In this method, the objective function *J* is minimized by introducing porous media into the flow regime. Comparing with the methods in the previous work, the following improvements have been made:

- Both Darcy's and Forchheimer's terms for porous medium flow have been taken into account.
- The effect of the porous media on the turbulent flow is considered according to our recent research [25].
- The sensitivity function with respect to the porosity is proposed.

A test case of application shows that a considerable reduction of the objective function can be obtained by this method. The transformation from the porous media to solid walls may have significant impact on the optimization results.

**Acknowledgments:**

The authors gratefully acknowledge the support of this study by the DFG (Deutsche Forschungsgemeinschaft).